\documentclass[a4paper]{jpconf}
\usepackage{graphicx}
\usepackage{epsfig,amsmath,amssymb} 
\usepackage[svgnames,table,hyperref]{xcolor}
\usepackage{graphics,color}
\usepackage{subfigure}

\newcommand{\Neel}{N\'{e}el }
\usepackage{cite}
\begin{document}
\title
{The ground-state magnetic ordering of the spin-1/2 frustrated $J_1$--$J_2$ XXZ model
on the square lattice}
\author{
R.~Darradi$^{1}$, J.~Richter$^{1}$ and J.~Schulenburg$^{2}$
}
\address
{
$^{1}$Institut f\"ur Theoretische Physik, Universit\"at Magdeburg, P.O. Box
4120,
D-39016 Magdeburg}
\address
{
$^{2}$Universit\"{a}tsrechenzentrum, Universit\"{a}t Magdeburg, P.O. Box
4120, 39016 Magdeburg
}
\ead{Rachid.Darradi@Physik.Uni-Magdeburg.DE}

\author{R.F.~Bishop$^{3}$ and P.H.Y.~Li$^{3}$}

\address
{
$^{3}$School of Physics and Astronomy, The University of Manchester, Manchester, M13 9PL, UK}

\begin{abstract}
Using the
coupled-cluster method for infinite lattices and the exact diagonalization
method
for finite lattices, 
we study the influence of an
exchange  anisotropy $\Delta$ on the ground-state phase diagram of the
spin-1/2 frustrated $J_1$--$J_2$ 
XXZ antiferromagnet on the square lattice. We find that increasing
$\Delta>1$ (i.e. an Ising type easy-axis anisotropy) 
as well as decreasing $\Delta < 1$ (i.e. an XY type easy-plane anisotropy)
both
lead to a monotonic shrinking of the parameter region 
of the magnetically disordered quantum phase. Finally, at $\Delta ^{c}
\approx 1.9$ 
this quantum phase disappears, whereas in the pure XY limit ($\Delta=0$) there is
still a narrow region around
$J_2 =0.5J_1$ where the quantum paramagnetic ground-state phase exists.
\end{abstract}

A canonical model to study the
interplay between frustration and quantum fluctuations in magnetic systems is 
the spin-1/2 Heisenberg antiferromagnet on the square lattice
{with nearest-neighbour (NN) coupling $J_1$ and frustrating
next-nearest-neighbour (NNN) coupling $J_1$ } 
($J_1$--$J_2$
model), see, 
e.g., Refs.~\cite{read91,schulz,Ig:1993,
richter94,capriotti03,singh03,Sir:2006,Darradi:2008}.
The recent syntheses of magnetic materials that can be well described 
by the $J_{1}$-$J_{2}$ model \cite{Mel:2000,Ro:2002}
has stimulated  further interest in the model.
For the isotropic spin-$1/2$ $J_1$--$J_2$ model 
there are two magnetically ordered ground state (GS) phases at small and at large 
$J_2$
separated by an intermediate  quantum  paramagnetic phase (QPP) without magnetic 
long-range order (LRO) in the region
$J_2^{c_1} \le J_2 \le J_2^{c_2}$,
where $J_2^{c_1} \approx 0.4J_1$ and  $ J_2^{c_2} \approx 0.6J_1$.
The GS at $J_2 < J_2^{c_1}$ exhibits N\'eel LRO.
The twofold degenerate GS at $J_2 > J_2^{c_2} $ shows so-called collinear 
magnetic LRO.
These two collinear GS's are characterized
by a parallel spin orientation of nearest neighbours in the vertical direction
and an antiparallel spin orientation of  nearest neighbours in the horizontal direction
[collinear-columnar (CC) state], and vice versa (collinear-row state).
The nature of the transition between the N\'eel phase and the QPP
as well as the properties of the QPP
are still under debate 
\cite{capriotti03,singh03,Sir:2006,Darradi:2008}.

Several extensions of the $J_1$--$J_2$ model have been studied recently, see,
e.g.,
Refs.~\cite{Be:1998,schmidt02,oitmaa04,Schm:2006,Ne:2003,Si:2004,Star:2004,Bi:2008_j1j2j3_spinHalf,
Ro:2004,Via:2007,Bi:2008_xxz_spinHalf}.
For instance, it was found that by increasing the spatial dimensionality from $D=2$ to $D=3$
the intermediate QPP disappears
\cite{Schm:2006,schmidt02,oitmaa04}.
{With respect to experimental realizations of the $J_1$--$J_2$
model an exchange anisotropy could be relevant. 
Surprisingly, only a few
papers so far have considered this issue. Some recent
papers have discussed the special cases where (i) only the NN coupling $J_1$ 
is anisotropic \cite{Via:2007} or, alternatively, where (ii) only the NNN coupling 
$J_2$
becomes anisotropic
\cite{Ro:2004}. In materials it seems to be more likely that both
couplings, $J_1$ and $J_2$, are anisotropic.
The corresponding model is the 
square-lattice spin-$\frac{1}{2}$  $J_{1}$-$J_{2}$ XXZ model
\begin{eqnarray} \label{Ham}
H = J_{1}\sum_{\langle i,j \rangle}(s^{x}_{i}s^{x}_{j}+s^{y}_{i}s^{y}_{j}+\Delta s^{z}_{i}s^{z}_{j}) + J_{2}\sum_{\langle\langle i,k \rangle\rangle}(s^{x}_{i}s^{x}_{k}+s^{y}_{i}s^{y}_{k}+\Delta s^{z}_{i}s^{z}_{k})\,,\label{H}
\end{eqnarray}
where the first sum 
runs over all NN and the second sum runs over all NNN pairs. 
To our best knowledge the only
study by other authors of such an anisotropic $J_1$--$J_2$ model with the same anisotropy
in the $J_1$ and the $J_2$ term has be performed  
by Benyoussef et al. \cite{Be:1998} using linear spin-wave
theory. Moreover these authors considered $\Delta \ge 1$ only.  
However, from early studies of Igarashi \cite{Ig:1993} of the isotropic 
$J_1$--$J_2$
model it is known 
that higher orders in the $1/s$ expansion  become large near the
quantum critical point (QCP), and hence results from the lowest order (i.e. linear) spin-wave theory
become  unreliable. }
As in our previous
work \cite{Schm:2006,Bi:2008_j1j2j3_spinHalf,Bi:2008_xxz_spinHalf,Darradi:2008} 
on $J_{1}$-$J_{2}$ models 
on the square lattice, we use here the coupled cluster method (CCM)
complemented by exact diagonalisation (ED) of a finite square lattice of
$N=36=
6 \times 6$ sites 
(imposing periodic boundary conditions)
to investigate the effect of exchange anisotropy.  
The CCM is an effective tool for studying highly frustrated 
quantum magnets
\cite{zeng98,Kr:2000,Bi:2008_j1j2j3_spinHalf,Schm:2006,Bi:2008_xxz_spinHalf,Darradi:2008},
where, e.g., the quantum Monte Carlo method is not applicable
due to the minus-sign problem.

For the CCM treatment of the model (\ref{Ham}), 
we use the classical GS (N\'eel 
at small $J_2$ and CC at large $J_2$) as reference state $\vert\Phi\rangle$.
Starting from these reference states the CCM employs the exponential
parametrization $|\Psi\rangle = e^S|\Phi\rangle$ of the quantum GS
$|\Psi\rangle $ 
where the correlation operator $S$  contains all possible multi-spin-flip
correlations present in the true GS. Naturally, $S$ has
to be approximated. 
We use the well-elaborated  CCM-LSUB$n$
approximation
\cite{zeng98,Kr:2000,Schm:2006,Bi:2008_j1j2j3_spinHalf,Bi:2008_xxz_spinHalf,Darradi:2008}
to calculate the GS energy per spin $E$ and the sublattice magnetization per
spin $M$.
Since the LSUB$n$ approximation becomes exact for $n \to \infty$,
it is useful to extrapolate the `raw' LSUB$n$ data to $n \to \infty$.
There are well-tested extrapolation formulas, namely 
$E(n) = a_0 + a_1(1/n)^2 + a_2(1/n)^4$
for the GS energy per spin \cite{Kr:2000,Bi:2000,Schm:2006,Bi:2008_j1j2j3_spinHalf,Bi:2008_xxz_spinHalf}
and 
$ M(n)=b_0+b_1(1/n)^{1/2}+b_2(1/n)^{3/2}$ for the 
sublattice magnetization
\cite{Bi:2008_j1j2j3_spinHalf,Bi:2008_xxz_spinHalf,Darradi:2008}.
We will not
present more details of the CCM, but rather refer, e.g.,  to 
Refs.~\cite{zeng98,Kr:2000,Bi:2000,Schm:2006,Darradi:2008}.

\begin{figure}
\begin{center}
\mbox{
\subfigure{\scalebox{0.54}
{\epsfig{file=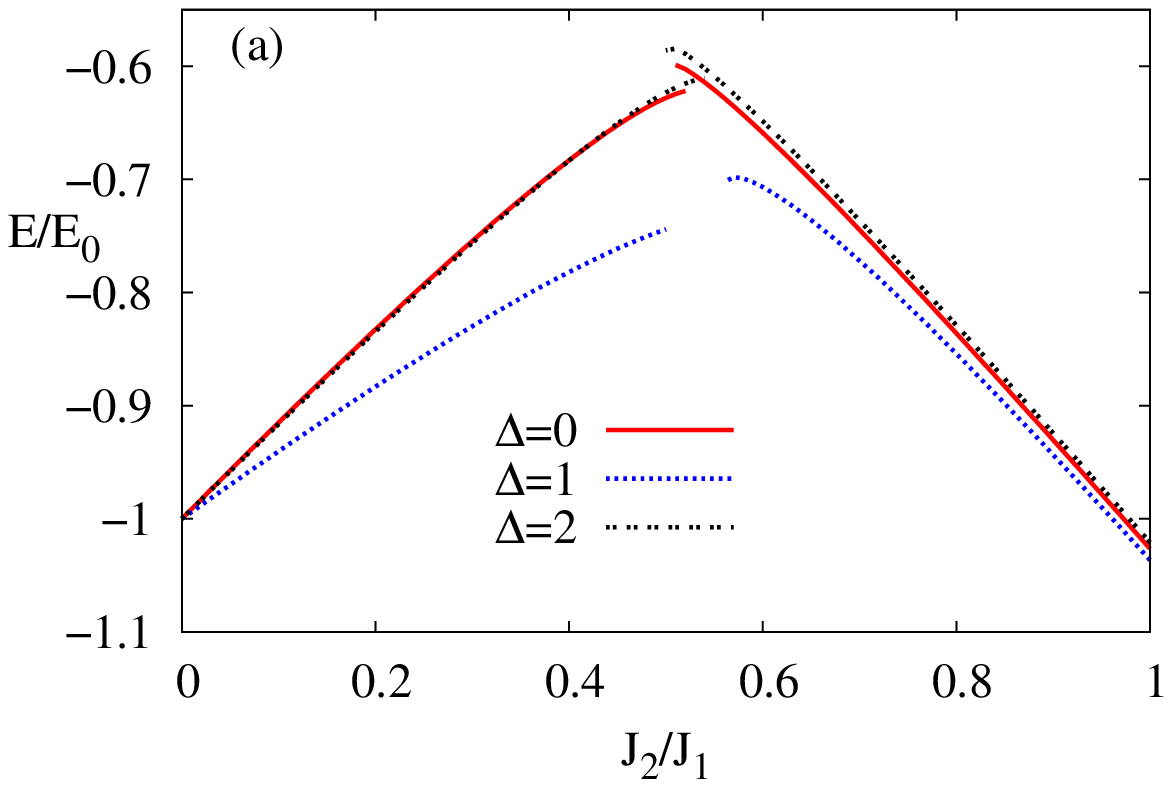,angle=0}}}
\subfigure{\scalebox{0.54}
{\epsfig{file=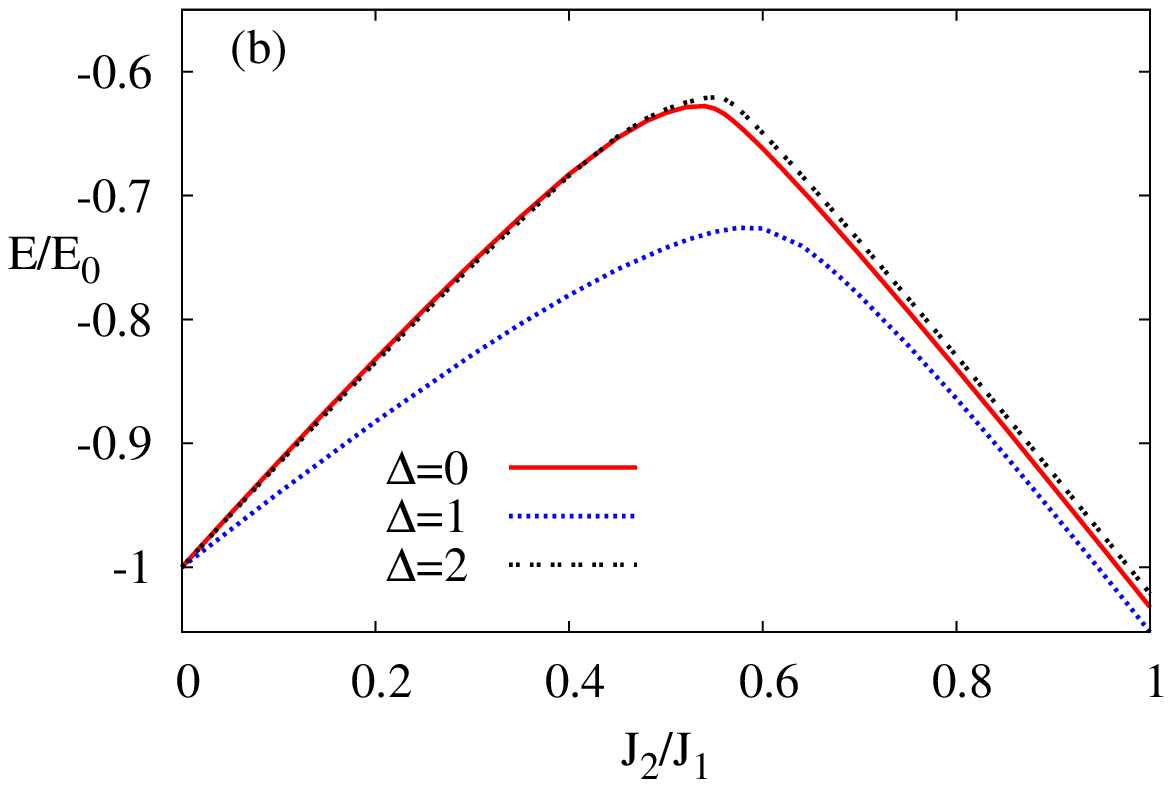,angle=0}}}
}
\caption{The GS energy per spin scaled by its value for $J_2=0$,
$E(J_2)/E(J_2=0)$, for anisotropies $\Delta= 0$ ($XY$), $\Delta=1$
(isotropic Heisenberg), $\Delta=2$ (Ising-type).    
(a) CCM: 
The LSUB$n$ results with $n=\{4,6,8,10\}$ are extrapolated to $n \rightarrow \infty$ using 
$E(n) = a_0 + a_1(1/n)^2 + a_2(1/n)^4$. (b) ED: $N=36$.}
\label{E}
\end{center}
\end{figure}

We start with the GS energy plotted in Fig.~\ref{E}
{for three
characterictic values of the anisotropy parameter $\Delta$}. As mentioned above 
we use for the CCM calculations  the N\'eel reference state  
at small $J_2$, but the  CC reference state at large
$J_2$. Hence, the CCM curves typically consists of two parts belonging to
N\'eel 
and CC reference states. 
The curves for $\Delta=1$ shown in Fig.~\ref{E}(a) 
illustrate clearly that
{there is a parameter region around
$J_2 = 0.5J_1$ where the
CCM equations using classical N\'eel and CC reference states do not have
real solutions. As a consequence, the    
corresponding pair of GS energy curves for the N\'{e}el and 
CC phases do not intersect one another.} This behaviour
yields 
preliminary evidence for the opening up of an intermediate phase between 
the N\'{e}el and CC phases. 
By contrast, for $\Delta=0$ and
$\Delta=2$  the corresponding pairs of GS energy curves almost cross one
another { giving preliminary evidence that for strong
anisotropy  a direct
first-order transition between both semiclassical magnetic phases may occur.}
From the ED data it is also evident, that the behaviour of the GS energy for
the isotropic model, i.e. $\Delta =1 $, differs from that for $\Delta=0$ and $\Delta=2$.

\begin{figure*}
\begin{center}
\mbox{
\subfigure{\label{CCM_M}\scalebox{0.28}
{\epsfig{file=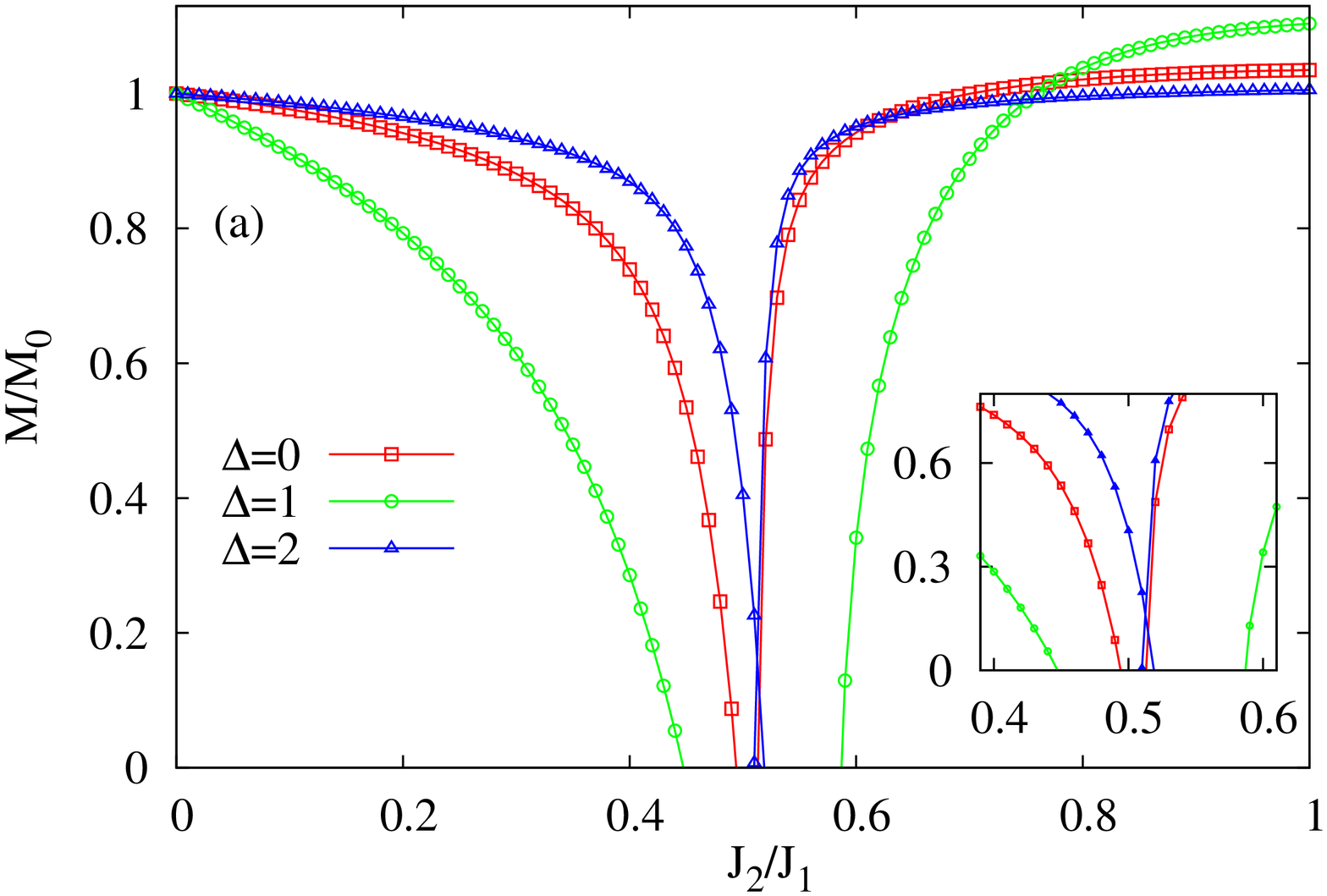,angle=270}}}
\subfigure{\label{ED_M}\scalebox{0.28}
{\epsfig{file=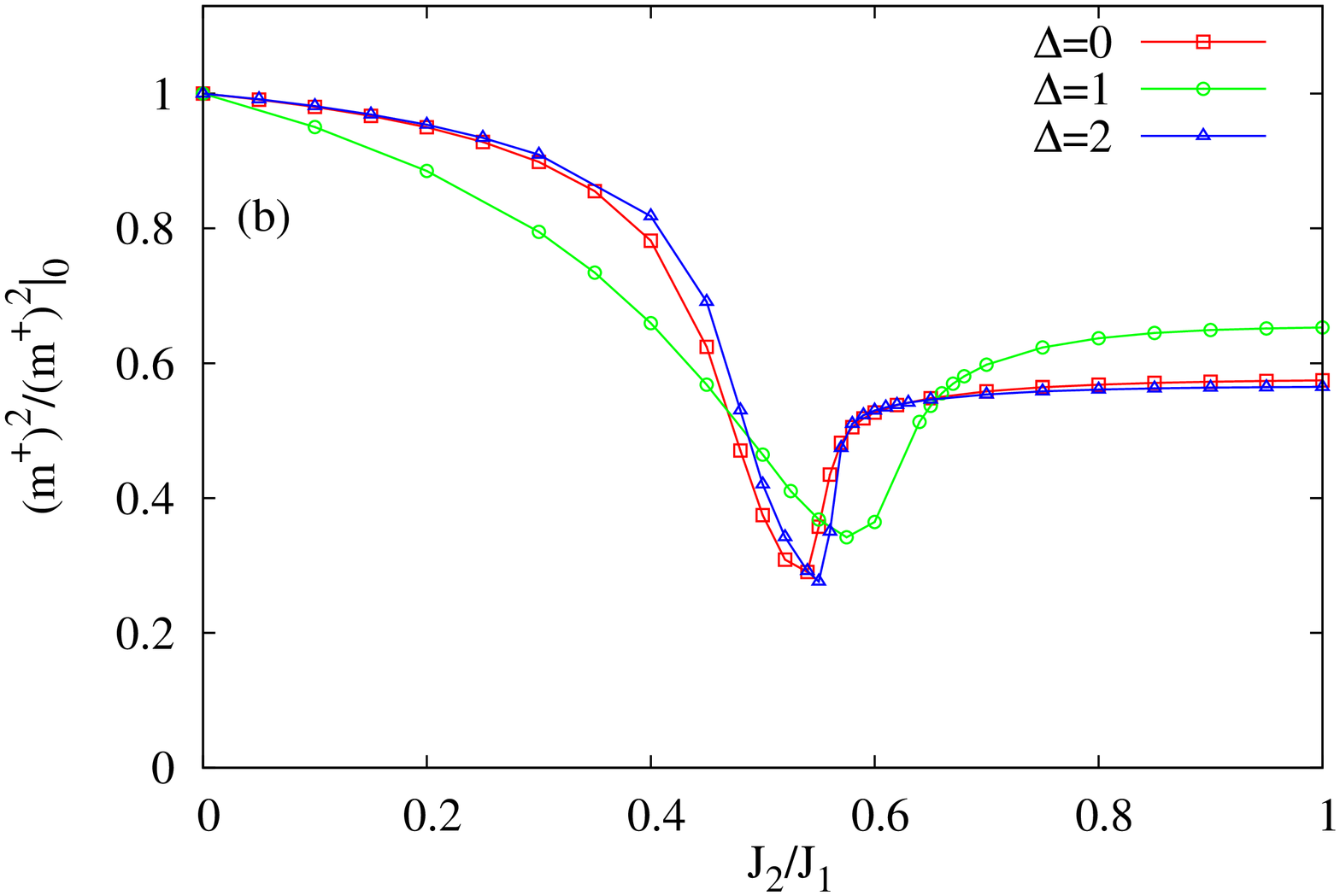,angle=270}}}
}
\caption{
Magnetic order parameter scaled by its value for $J_2=0$
for anisotropies $\Delta= 0$ ($XY$), $\Delta=1$
(isotropic Heisenberg), $\Delta=2$ (Ising-type).
(a) CCM: Sublattice magnetization $M(J_2)/M(J_2=0)$.  
The LSUB$n$ results for $M$ with $n=\{4,6,8,10\}$ are extrapolated to $n \rightarrow
\infty$ using
$ M(n)=b_0+b_1(1/n)^{1/2}+b_2(1/n)^{3/2}$. {The inset shows the
parameter region around $J_2 =0.5J_1$ with an enlarged scale.}
(b) ED: Square of the order parameter
$(m^+)^2(J_2)/(m^+)^2(J_2=0)$  for $N=36$.}
\label{M}
\end{center}
\end{figure*}

Next we present in Fig.~\ref{M} the sublattice magnetization  $M$ 
calculated by the CCM for
$N\to \infty$ and 
the  square of an order parameter defined as 
$(m^+)^2 =
\frac{1}{N^2}\sum^N_{i,j}\vert \langle{\bf s}_i \cdot {\bf s}_j \rangle \vert $ \cite{Ri:2004} 
calculated 
by ED for  $N=36$. {More data for $M$ can be found in
Ref.~\cite{Bi:2008_xxz_spinHalf}.}
While $M$ is finite in the magnetically ordered phases
but vanishes in the intermediate QPP, we have always finite
values for $(m^+)^2$ for finite $N=36$.
{We use the CCM results for $M$ to define the QCP's
$(J_2^{c_1}, \Delta_{c_1})$ and $(J_2^{c_2}, \Delta_{c_2})$ as that
points where the magnetic order
parameter [\Neel at $(J_2^{c_1}, \Delta_{c_1})$ and CC at $(J_2^{c_2},
\Delta_{c_2})$]  becomes zero. 
From Fig.~\ref{M} it is obvious
that the intermediate QPP is largest for $\Delta = 1$
(the CCM estimates for $J_2^{c_1}$ and $J_2^{c_2}$ are $J_2^{c_1}\approx 0.44 \ldots 0.45J_1$
and
$J_2^{c_2}\approx 0.58 \ldots 0.59J_1$ for $\Delta =1$, 
cf. Refs.~\cite{Bi:2008_xxz_spinHalf,Darradi:2008}).
Both types of anisotropy lead to a stabilization of magnetic
LRO.  
The ED data for $(m^+)^2$ support these findings.  From Fig.~\ref{M}(b)  
it can be seen that the parameter region of small values of  $(m^+)^2$ around 
$J_2=0.5J_1$ is significantly broader for $\Delta =1$ than for $\Delta =2$ and $\Delta
=0$. 
From the inset of Fig.~\ref{M}(a) it is also obvious that 
the pair 
of CCM 
order-parameter curves (N\'{e}el
and CC) for $\Delta=2$
intersect one another at a  value $M
\geq 0$. This behaviour can be interpreted as an indication of a 
first-order transition between the magnetically ordered N\'{e}el and CC
phases located at the crossing point, for a detailed discussion of this
issue see also
Refs.~\cite{Bi:2008_xxz_spinHalf,Schm:2006}. 
} 

\begin{figure}[t]
\begin{center}
\mbox{
\subfigure{\scalebox{0.27}
{\epsfig{file=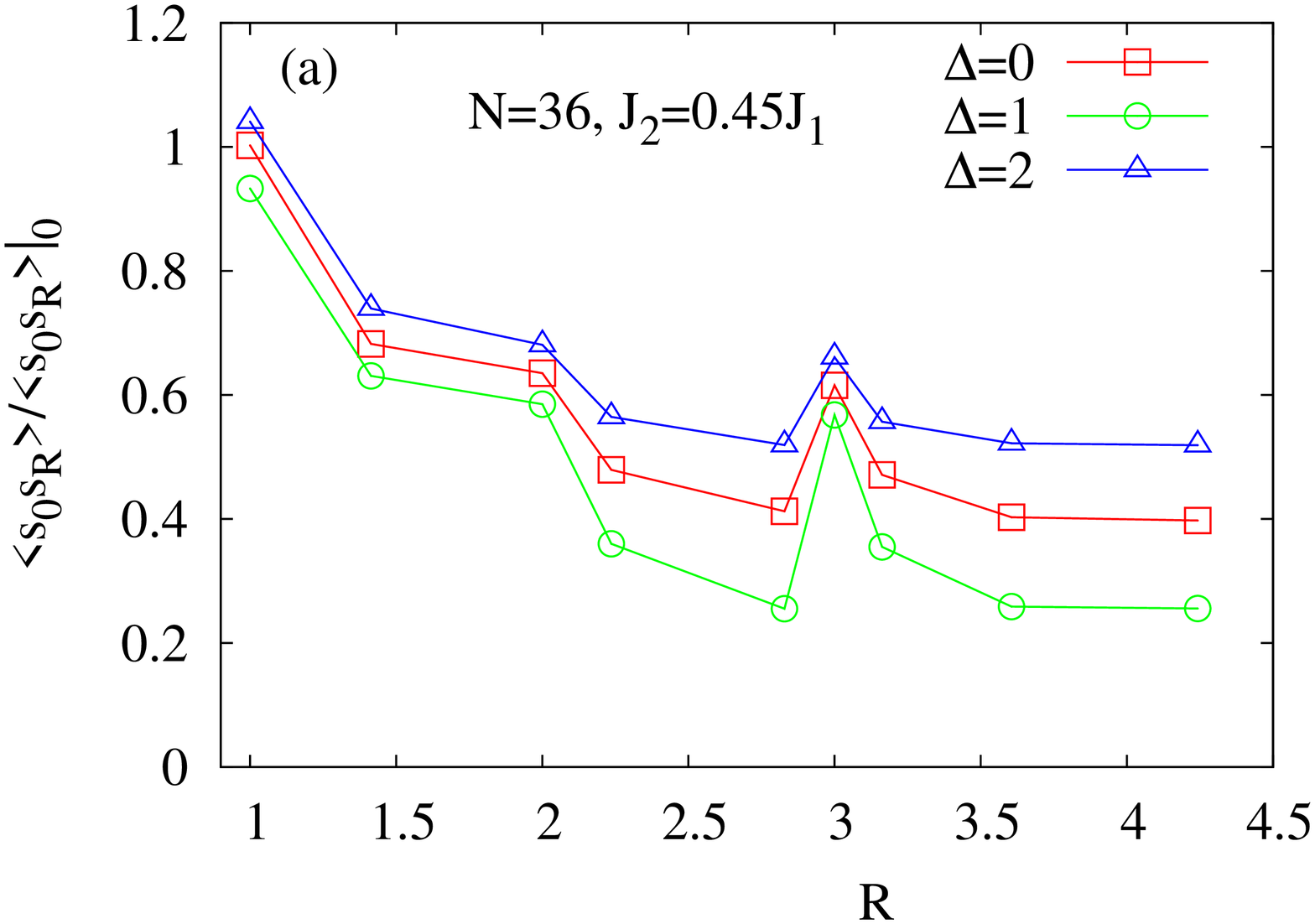,angle=270}}}
\subfigure{\scalebox{0.28}
{\epsfig{file=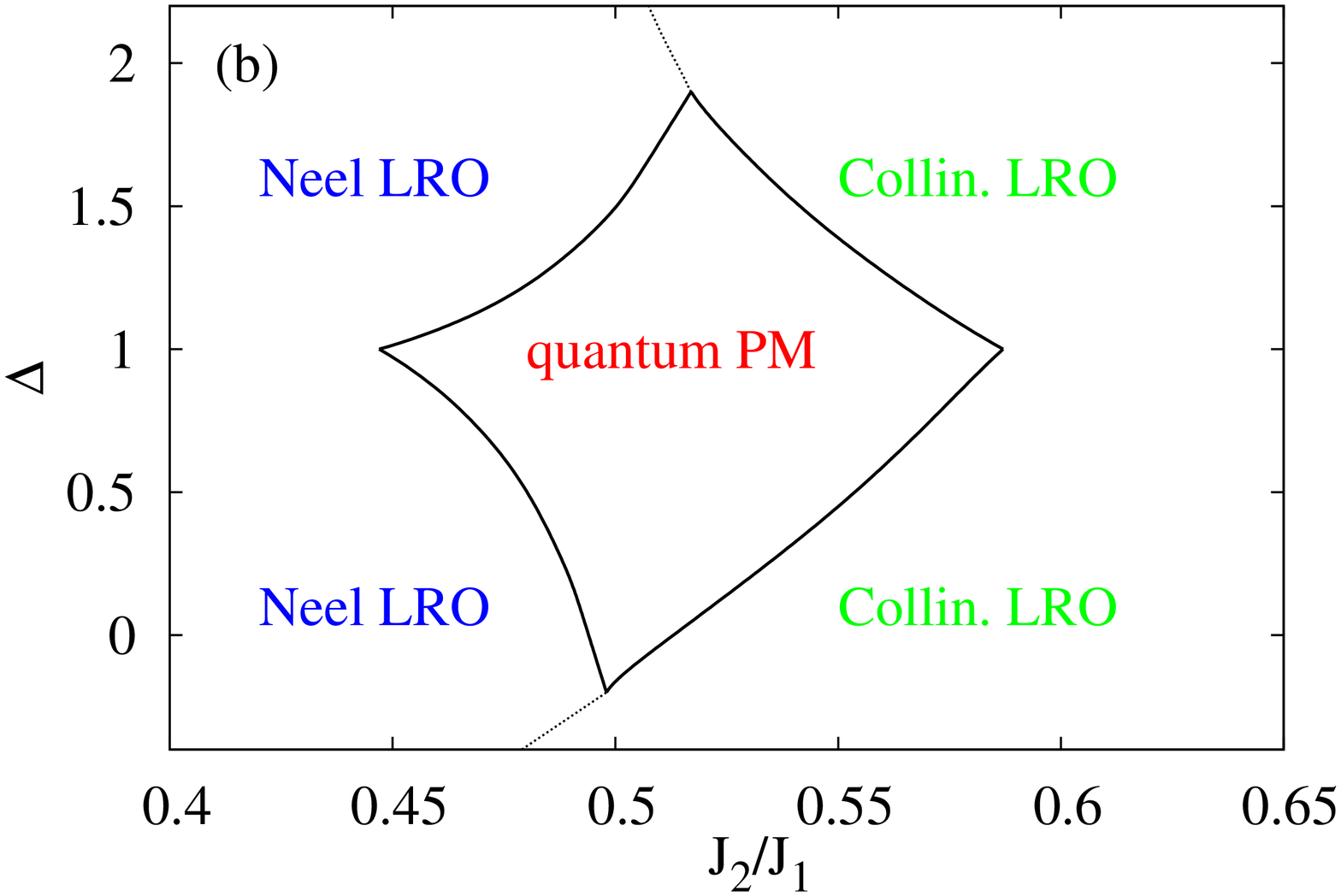,angle=270}}}
}
\caption{(a): 
ED results for the spin correlation function scaled by its value 
for $J_2=0$, 
$\langle {\bf s}_0 \cdot {\bf s}_R\rangle (J_2)/\langle {\bf s}_0 \cdot {\bf
s}_R\rangle (J_2=0)$ versus separation $R$ 
for $\Delta= 0$, $1$, and  $2$ and $J_2=0.45J_1$.
(b) GS phase diagram of the spin-1/2 $J_1$--$J_2$ XXZ model on the
square lattice calculated by the CCM.
}
\label{phase}
\end{center}
\end{figure}

{The influence of the anisotropy on the 
correlator $\langle {\bf s}_0 \cdot {\bf s}_R\rangle$ is illustrated in Fig.~\ref{phase}(a)
for $J_2=0.45J_1$, i.e. near 
the QCP
$J_2^{c_1}$ where the \Neel  LRO breaks down for $\Delta=1$.}
We see that $\langle {\bf s}_0 \cdot {\bf s}_R\rangle$
decays most rapidly for $\Delta=1$. For the
largest separation $R_{\mbox{\small  max}}=\sqrt{18}$ present 
for $N=36$ sites, the
correlator 
$\langle {\bf s}_0 \cdot {\bf s}_{R_{\mbox{\small max}}}\rangle$ for
$J_2=0.45J_1$ is reduced by
frustration by a
factor of $0.25$
for $\Delta =1$, whereas the corresponding factor is only 
$0.52$ ($0.40$) for $\Delta=2$
($\Delta=0$).

{Finally, we summarize our findings on the GS magnetic
ordering in Fig.~\ref{phase}(b) where the GS phase diagram is shown.
The solid lines in Fig.~\ref{phase}(b) represent those
points in the ($\Delta, J_2$) parameter space  where the \Neel and the CC order
parameter calculated by the CCM vanish.}
Increasing the anisotropy leads to a monotonic shrinking of the
region of the QPP.
For the  easy-axis anisotropy 
all three phases meet at a         
 quantum triple point  at 
($\Delta ^{c}  \approx 1.9, J_{2}^{c} \approx 0.52$), i.e. the QPP
disappears completely for $\Delta \gtrsim 1.9$.
Similarly, for the case of the easy-plane anisotropy a second  quantum triple point occurs at 
($\Delta ^{c} \approx -0.1 , J_{2}^{c} \approx 0.50$).
{Outside the area hemmed by the solid lines 
there is a direct first-order transition between the \Neel and the CC phase,
as indicated by the dotted lines.}


\end{document}